\documentstyle[aps,prl,psfig,multicol]{revtex}
\begin{document}

\title{Cyclotron resonance in a two-dimensional
electron gas with long-range randomness}

\author{M. M. Fogler}

\address{School of Natural Sciences, Institute for Advanced Study,
Olden Lane, Princeton, NJ 08540}

\author{B. I. Shklovskii}

\address{Theoretical Physics Institute, University of Minnesota,
116 Church St. Southeast, Minneapolis, Minnesota 55455}

\date{
\today\
}

\maketitle

\raisebox{125pt}[0pt][0pt]{\mbox{\hspace{5.0in}Preprint IASSNS-HEP-98/001}}

\vspace{-0.3in}

\begin{abstract}

We show that the the cyclotron resonance in a two-dimensional electron
gas has non-trivial properties if the correlation length of the disorder
is larger than the de Broglie wavelength: (a) the lineshape assumes
three different forms in strong, intermediate, and weak magnetic fields
(b) at the transition from the intermediate to the weak fields the
linewidth suddenly collapses due to an explosive growth in the fraction
of electrons with a diffusive-type dynamics.

\end{abstract}
\pacs{PACS numbers: 73.40.Hm}

\begin{multicols}{2}
%\narrowtext

The cyclotron resonance (CR) is one of the basic tools for studying the
electronic properties of physical systems in an external magnetic field.  A
very interesting example of such a system is a two-dimensional electron gas
(2DEG), which exhibits the quantized Hall effect if the magnetic field is
sufficiently strong. The CR can be studied by measuring
the transmission of an electromagnetic signal of some frequency $\omega$
through the 2DEG.  If the absorption is weak, the change in the
transmission is proportional to the real part of the dynamical conductivity
$\sigma_{xx}(\omega)$, which is the average over the active and inactive
circular polarizations, ${\rm Re}\,\sigma_{xx}(\omega) = [{\rm
Re}\,\sigma_+(\omega) + {\rm Re}\,\sigma_-(\omega)] / 2$ (for linearly
polarized signals). The active polarization's contribution ${\rm
Re}\,\sigma_+(\omega)$ has a peak at $\omega$ close to the cyclotron
frequency $\omega_c$ of the external magnetic field. 
The disorder-related zero-temperature width and shape of this peak are
the subjects of this Letter.

\begin{figure}
\centerline{
\psfig{file=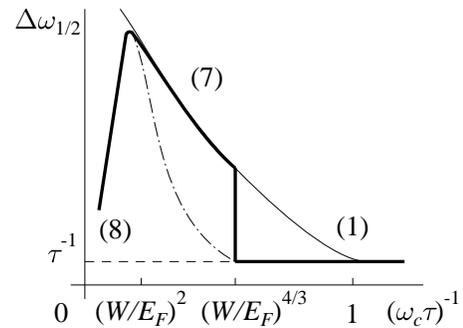,width=2.3in,%
bbllx=170pt,bblly=285pt,bburx=436pt,bbury=480pt}}
\vspace{0.1in}
\setlength{\columnwidth}{3.2in}
\centerline{\caption{
Dependence of the CR linewidth on $(\omega_c\tau)^{-1}$, the quantity
inversely proportional to the magnetic field. Thick line: our results for
the long-range potential with given $W/E_F$ and $\tau$ [for the case
$k_F d \gg (E_F/W)^{2/3}$]. Labels (\protect\ref{Drude}),
(\protect\ref{delta_omega_strong}), and
(\protect\ref{delta_omega_intermediate}) correspond to the equation
numbers. Thin line: short-range potential with the same $\tau$.
Dash-dotted line: the width $\nu$ of an additional structure near the
very resonance (see text).
\label{linewidth}
}}
\end{figure}

The theoretical study of the CR in the 2DEG has been initiated by
Ando~\cite{Ando}. Although many pieces of the complete picture has been
put in place by him and later by other authors~\cite{Prasad,Bychkov},
the consistent description of this phenomenon exists only for the case
where the correlation length $d$ of the random potential $U(x, y)$
acting on the electrons in the disordered 2DEG is smaller than the de
Broglie wavelength $k_F^{-1}$ ($k_F$ is the Fermi wavevector in zero
magnetic field). In this case the effect of the random potential is
described by a single quantity, the transport (or momentum relaxation)
time $\tau$. It was understood, however, that the conventional
Drude-Lorentz formula
\begin{equation}
{\rm Re}\,\sigma_+(\omega) =
\frac{\sigma_0}{1 + (\omega - \omega_c)^2 \tau^2}
\label{Drude}
\end{equation}
($\sigma_0$ is the zero-field conductivity), which gives the Lorentzian peak
with the half width at half maximum (HWHM) of $\tau^{-1}$, applies only in
the uninteresting case $\omega_c\tau \ll 1$. In the other limit
($\omega_c\tau \gg 1$) the CR lineshape is non-Lorentzian and has a
much larger width~\cite{Ando}
\begin{equation}
             \Delta\omega_{1/2} = 0.73\, \sqrt{\omega_c / \tau}
\label{SCBA}
\end{equation}
due to the formation of discrete Landau levels.
This behavior of $\Delta\omega_{1/2}$ is illustrated by the thin line in
Fig.~\ref{linewidth}.

%
% FIG. 1
%

Here and below $\Delta\omega_{1/2}$ is the {\it median\/} width
defined by
\[
\int_0^{\Delta\omega_{1/2}}\! d \omega\,
{\rm Re}\,\sigma_+(\omega + \omega_c) = \frac12
\int_0^{\infty}\! d \omega\,{\rm Re}\,\sigma_+(\omega + \omega_c).
\]
Using the median width instead of the conventional HWHM is more adequate
because the CR lineshape can be rather intricate for a long-range random
potential.

For simplicity, we will consider a model of a Gaussian random potential
whose correlator decays sufficiently fast at $r > d$ and does not
possess any other characteristic scales besides $d$. The rms amplitude
of $U$ will be denoted by $W$. Until the very end we will assume that
$k_F d \gg (E_F / W)^{2/3}$, where $E_F$ is the Fermi energy. As one can
see from Fig.~\ref{linewidth} illustrating our results for this case,
the dependence of $\Delta\omega_{1/2}$ on the magnetic field is
non-monotonic. Even more remarkable, $\Delta\omega_{1/2}$ exhibits a
rapid {\it collapse\/} to its classical value of $\tau^{-1}$ in the
vicinity of the point $\omega_c \tau \sim (E_F / W)^{2/3} \gg 1$.
%Note that there exists a wide range of magnetic
%fields where both Eq.~(\ref{Drude}) is valid and $\omega_c \tau \gg 1$.
The derivation of these results is based on the picture of the
``classical localization''~\onlinecite{Fogler_97}, which we will now
discuss.

If the random potential $U$ is smooth and the time scale on which
we study the motion of an electron is not too long, then it can be
described classically, as a motion of a single particle with energy
$E_F$. (We neglect the interaction). If the magnetic field is not too
low, the motion can be decomposed into a fast cyclotron gyration and a
slow drift of the guiding center $\bbox{\rho} = (\rho_x, \rho_y)$ of the
cyclotron orbit. For magnetic field in the negative $z$-direction,
$\rho_x = x + (v_y / \omega_c)$ and $\rho_y = y - (v_x / \omega_c)$,
where $x$ and $y$ are the coordinates of the electron and $\bbox{v} =
-v_F\,(\sin\theta, \cos\theta)$ is its velocity. We will call the
magnetic field {\it strong\/} if the cyclotron radius $R_c = v_F /
\omega_c$ is smaller than $d$. It is usually assumed that in such strong
fields the guiding center drifts along a level line
$U(\bbox{\rho}) = \text{const}$ of the random potential, which is
typically a closed loop of size $d$.
Recently it has been realized~\cite{Fogler_97,Laikhtman} that the drift
approximation is also valid in the {\it intermediate field\/} regime $1
< R_c / d < (E_F / W)^{2/3}$ [the same as $(E_F / W)^{4/3} <
\omega_c\tau < (E_F / W)^2$ because $\tau \sim (d / v_F) (E_F / W)^2$].
The point is that in this regime the cyclotron gyration is still
sufficiently fast, so
that the guiding center remains practically ``frozen'' during one
cyclotron period. Therefore, the guiding center motion is determined by
$U_0$, the random potential {\it averaged\/} over the cyclotron orbit,
\[
U_0(\bbox{\rho}, R_c) = \int_0^{2\pi}\!\frac{d \theta}{2 \pi}\,
  U(\rho_x + R_c \cos \theta, \rho_y + R_c \sin \theta).
\]
The guiding center is still bound to one of the level lines, but those
are the level lines of $U_0$, not $U$~\cite{Fogler_97}.  The rms
amplitude $W_0$ of the potential $U_0$ is by a factor of $\sqrt{d /
R_c}$ smaller than $W$ due to the averaging, but the correlation length
is the same; therefore, a typical level line of $U_0$ is still a loop of
size $d$. The frequency $\omega_d$ of the guiding center motion along
such a loop (the drift frequency) is given by $W_0 / (m \omega_c d^2)$,
which can be cast into the form
\begin{equation}
   \omega_d = (R_c / d) \sqrt{\omega_c / \tau},
   \quad R_c \gg d.
\label{omega_d_intermediate}
\end{equation}
Obviously, the electrons on the periodic orbits are
(classically) localized and do not participate in the DC transport.
However, a more accurate analysis~\cite{Fogler_97} reveals that a
very small fraction of the order of
$e^{-(\omega_c/\omega_d)^{2/3}}$ of the trajectories remains
delocalized. Such trajectories form a stochastic web in the
vicinity of the percolation contour. The stochastic
web rapidly grows with decreasing magnetic field and turns into
a stochastic sea at
\begin{equation}
                       R_c / d = (E_F/W)^{2/3},
\label{boundary}
\end{equation}
where $\omega_d = \omega_c$. In even lower magnetic fields (the {\it
weak field} regime) the stochastic sea spans almost the entire phase
space. Correspondingly, the static conductivity is exponentially small,
$\sigma_{xx}(0) \propto e^{-(\omega_c/\omega_d)^{2/3}}$ when $\omega_d
\ll \omega_c$ (the strong and the intermediate field regimes), rapidly
blows up near $\omega_d = \omega_c$ point (the boundary of the
intermediate and the weak field regimes), and finally crosses over to
the Drude-Lorentz formula~(\ref{Drude}) in weak
fields~\cite{Comment_on_DC}.

Our goal is to show that the {\it
dynamical\/} conductivity also exhibits a rapid change near the
$\omega_d = \omega_c$ point: the aforementioned collapse of the CR
linewidth. This can be done
on the basis of the classical formula for $\sigma_+(\omega)$,
\begin{equation}
\sigma_+(\omega) = \frac{\sigma_0}{\tau} \!\int_0^\infty\! d t\,
\left\langle e^{i\omega t - i[\theta(t) - i\theta(0)]} \right\rangle.
\label{Kubo}
\end{equation}
Suppose that $\Delta\omega \equiv \omega - \omega_c$ is smaller than
$\omega_c$ by absolute value, then the full equation of motion for
$\theta$ can be replaced by its average over the
cyclotron period,
\begin{equation}
 \frac{d \theta}{d t} = \omega_c + \delta\omega,\quad
 \delta\omega = \left.\frac{1}{m\omega_c R}
        \frac{\partial}{\partial R}U_0(\bbox{\rho}, R)\right|_{R = R_c}
\label{theta}
\end{equation}
where $\delta\omega$ is the {\it local correction\/} to the gyration
frequency of the velocity vector.  The origin of such a correction is
quite clear.  Indeed, the random potential creates an additional
centripetal force, which is either parallel to the Lorentz force and
thus speeds up the cyclotron gyration, or anti-parallel, which slows it
down.

In the strong and intermediate field regimes most of the guiding centers
are localized on the periodic orbits of small size $\lesssim d$. The
correction $\delta\omega$ does not vary much on this length scale. In
fact, to find $\Delta\omega_{1/2}$ up to a numerical factor, we can use
the approximation $\delta\omega = \text{const}$. Substituting this into
Eqs.~(\ref{Kubo}) and (\ref{theta}), we find that ${\rm
Re}\,\sigma_+(\omega)$ has the form of a Gaussian peak of width
$[\delta\omega]_{\rm rms}$, and so
\begin{eqnarray}
\Delta\omega_{1/2} \sim
[\delta\omega]_{\rm rms} &=& \text{const}\frac{W}{m \omega_c d^2},
\quad R_c \ll d,
\label{delta_omega_strong}\\
\mbox{} &=& \sqrt{\omega_c / (\pi\tau)}, \quad R_c \gg d.
\label{delta_omega_intermediate}
\end{eqnarray}
The broadening of the CR line is of the inhomogeneous type, and comes
from the places where $\Delta\omega =
\delta\omega$~\cite{Comment_on_shape}.
Equations~(\ref{delta_omega_strong}) and (\ref{delta_omega_intermediate})
show that $\Delta\omega_{1/2}$ linearly increases as a function of
$(\omega_c\tau)^{-1}$ in the strong-field regime, reaches its maximum at
$R_c \sim d$, and then decreases in the intermediate fields, see
Fig.~\ref{linewidth}.

The situation is different in the weak fields where most of the
trajectories are extended and ergodic. In this case the broadening is of
the homogeneous type and is much smaller than $[\delta\omega]_{\rm rms}$
because of the motional narrowing. The crossover between the two types
of broadening occurs at the low-field end of the intermediate-field
regime, which we study below. (However, a broader physical discussion will
be resumed in the concluding remarks).

According to Eqs.~(\ref{omega_d_intermediate}) and
(\ref{delta_omega_intermediate}), $[\delta\omega]_{\rm rms} \ll
\omega_d$ in this regime. Because of this inequality, the orbits of
{\it low\/} frequency $\Omega \ll \omega_d$ whose perimeter length
is typically $(\omega_d / \Omega) d$ turn out to be
important. To show this we present $\sigma_+(\omega)$ as follows,
\begin{eqnarray}
\sigma_+(\omega) &=& \frac{\sigma_0}{\tau} \int_0^\infty\!\!d \Omega\,
f(\Omega)\, I(\Omega, \Delta\omega),
\label{sigma_plus_2}\\
I(\Omega, \Delta\omega) &=& \int_0^\infty\!\! d t
\left\langle\!\exp\!\left[i \Delta\omega t
- i\int_0^t\!\! d t' \delta\omega(t')\right]
\right\rangle_\Omega,
\label{I}
\end{eqnarray}
where $\langle\ldots\rangle_\Omega$ denotes the average over the orbits
of frequency $\Omega$, and $f(\Omega)$ is the probability density of
finding an electron with energy $E_F$ on such an orbit (with the
convention $\Omega = +0$ for the unbound trajectories of the stochastic
web). The correlations in $\delta\omega(\bbox{\rho})$ decay sufficiently
fast ($\sim \rho^{-3}$) with distance; hence, the high-order
correlators of $\delta\omega$ can be neglected, and we arrive at
\[
I \simeq \int_0^\infty\!\! d t\,
\exp\!\left[i\Delta\omega t - \frac12\int_0^t\! d t_1\!\int_0^t\! d t_2\,
\langle \delta\omega(t_1) \delta\omega(t_2) \rangle_\Omega\right].
\]
For the same reason we can approximate the second-order correlator by
the sum of isolated short pulses,
\begin{equation}
\langle \delta\omega(t_1) \delta\omega(t_2) \rangle_\Omega =
\sum_{n = -\infty}^\infty P(t_1 - t_2 - \frac{2 \pi n}{\Omega}),
\label{delta_omega_corr}
\end{equation}
where $P(0) = \langle\delta\omega^2\rangle$ and $P(t) \ll P(0)$ for $|t|
\gg \omega_d^{-1}$. In fact, if both $\Omega$ and $|\Delta\omega|$ are
much smaller than $\omega_d$, the actual functional form of the pulses
is unimportant, and we can approximate $P(t)$ by $2 \nu\,\delta(t)$,
where
\begin{equation}
\nu \sim \langle\delta\omega^2\rangle / \omega_d
    \sim \omega_c / (\omega_d\tau).
\label{nu}
\end{equation}
After such approximations and some algebra we find:
\begin{eqnarray}
{\rm Re}\, I &=& \frac{\pi}{\sqrt{2 \Omega \nu}}
\sum_{n = -\infty}^\infty c_n\,
\exp\!\left[-\frac{\pi (\Delta\omega - n\Omega)^2}{2 \Omega \nu}
\right],
\label{I_sum}\\
c_n &=& \int_0^1\! d x\,
\cos(\pi n x)\,\exp\!\left[-\frac{\pi \nu (2 x - x^2)}{2 \Omega}\right].
\end{eqnarray}
If $\Omega \gg \nu$, then $c_0 \simeq 1$ and $c_n \simeq \nu / (\pi n^2
\Omega)$, $n \neq 0$, so that ${\rm Re}\, I(\Omega, \Delta\omega)$ as a
function of $\Delta\omega$ is the sum of narrow Gaussians centered at
equidistant points $\Delta\omega = n \Omega$.
In the opposite limit the Gaussians merge
into a single Lorentzian,
\begin{equation}
{\rm Re}\, I(\Omega, \Delta\omega) = \frac{\nu}{(\Delta\omega)^2 + \nu^2},
\quad \Omega \ll \nu,\:\:|\Delta\omega| \ll \omega_d.
\label{I_Lorentzian}
\end{equation}
There is a simple way to understand this equation. As discussed above,
the random potential modifies the local gyration frequency of the
velocity vector by $\delta\omega$. The sign of the correction changes
randomly each time the guiding center travels the distance $\sim d$.
This causes the angular diffusion of the velocity vector in addition to
the regular precession with frequency $\omega_c$ and therefore, the
momentum relaxation. The time $\omega_d / \langle \delta\omega^2 \rangle
= \nu^{-1}$ plays the role of the effective transport time, and so
Eq.~(\ref{I_Lorentzian}) is expected by analogy with Eq.~(\ref{Drude}).

To finish the calculation of $\sigma_+(\omega)$ we need to know
$f(\Omega)$. It is determined by the statistical properties of
the potential $U$, {\it viz.\/}, by the exponent $H$ in the equation
\[
\int_q^{2 q}\! d q q J^2_0(q R_c) \tilde{C}(q) \propto q^{-2 H},
\]
where $J_0$ is the Bessel function and $\tilde{C}(q)$ is the
Fourier transform of $C(\bbox{r}) \equiv \langle U(0) U(\bbox{r})\rangle$. 
In a practically relevant case where the potential is created by
randomly positioned donors confined to a narrow layer separated by the
distance $d$ from the 2DEG, $\tilde{C}(q) = \tilde{C}(0) e^{-2 q d}$, so
that $H = -1$ if $q \ll R_c^{-1}$ and $H = -1 / 2$ if
$R_c^{-1} \lesssim q \lesssim d^{-1}$. In both cases
\begin{mathletters}
\label{f}
\begin{eqnarray}
f(\Omega) &\simeq& s \Omega^{-1}\,(\Omega / \omega_d)^s,
\quad \Omega_{\rm web} < \Omega < \omega_d,\\
 &\simeq& (\Omega_{\rm web}/\omega_d)^s\,\delta(\Omega - 0)\quad
\text{at other }\Omega,
\end{eqnarray}
\end{mathletters}
where $s$ is very close to $1/2$ [see Isichenko~\onlinecite{Isichenko}
for details; $\Omega f(\Omega)$ corresponds to his $F(\Omega)$]. The
quantity $\Omega_{\rm web} \sim \omega_d e^{-(\omega_c /
\omega_d)^{2/3}}$ is the {\it drift\/} frequency of such orbits within
the stochastic web that give the dominant contribution~\cite{Fogler_97}
to the DC transport.

\begin{figure}
\centerline{
\psfig{file=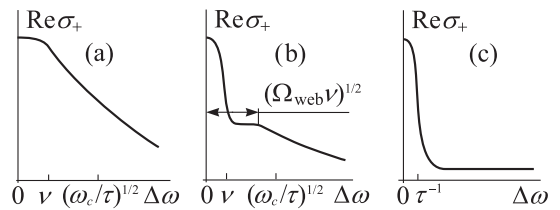,width=2.8in,%
bbllx=85pt,bblly=340pt,bburx=574pt,bbury=522pt}}
\vspace{0.1in}
\setlength{\columnwidth}{3.2in}
\centerline{\caption{The CR lineshape (schematically).
(a) before the collapse
(b) in the course of the collapse
(c) immediately after the collapse.
\label{lineshape}
}}
\end{figure}

We can now substitute Eqs.~(\ref{I_sum}-\ref{f}) into
Eq.~(\ref{sigma_plus_2}). Since $s < 1$, only the $n = 0$ term in
Eq.~(\ref{I_sum}) is important for $\nu \lesssim \Omega \lesssim
\omega_d$. Also, it is safe to assume that $\omega_d / \nu \ll e^{2 / (2
s - 1)}$~\cite{Comment_on_s}. With this in mind, we obtain ${\rm
Re}\,\sigma_+(\omega) = (\sigma_0 / \nu\tau)\, [S_1(\omega) +
S_2(\omega)]$, where
\begin{eqnarray*}
S_1(\omega) &\approx&
\sqrt{\frac{\Omega_c}{\omega_d}}\,
%\left(\frac{\sigma_0}{\nu\tau}\right)
\frac{1}{1 + (\Delta\omega / \nu)^2},\:\:|\Delta\omega|\lesssim\omega_d,
%\label{S_1}
\\
S_2(\omega) &\approx&
\sqrt{\frac{\pi^2 \nu}{8\omega_d}}\,
%\left(\frac{\sigma_0}{\nu\tau}\right)
\ln\left[
\frac{\omega_d\,\nu}{\max\,\{(\Delta\omega)^2, \Omega_c\nu\}}
\right],\:\: |\Delta\omega|\lesssim\sqrt{\omega_d \nu},
%\label{S_2}
\end{eqnarray*}
$\Omega_c = \max\,\{\Omega_{\rm web},\, \nu\}$. It is easy to see that
as long as $\Omega_{\rm web} \leq \nu \ll \omega_d$,
$\Delta\omega_{1/2}$ is determined by $S_2(\omega)$, and thus is of the
order of $\sqrt{\omega_d \nu} \sim \sqrt{\omega_c / \tau}$, in
accordance with Eq.~(\ref{delta_omega_intermediate}). The CR lineshape
for this case is depicted in Fig.~\ref{lineshape}a.

%
% FIG. 2
%

Let us now discuss the aforementioned collapse of the CR linewidth.
$\Omega_{\rm web}$ grows exponentially as the magnetic field decreases
and eventually becomes larger than $\nu$.  From this moment, the maximum
value of $S_2$ rapidly decreases and that of $S_1$ does the opposite.
It gives rise to a narrow Lorentzian peak at the center of the CR line
(see Fig.~\ref{lineshape}b).  At the $\omega_d = \omega_c$ point where
$\Omega_{\rm web} \sim \omega_d$, it is this peak that determines
$\Delta\omega_{1/2}$.  Thus, $\Delta\omega_{1/2}$ drops to the value of
$\nu = \omega_c/(\omega_d\tau) = \tau^{-1}$ (Fig.~\ref{lineshape}c).
From this point on, i.e., in the weak-field regime, Eq.~(\ref{Drude})
applies.

Several comments are in order here. Formula~(\ref{delta_omega_strong})
for the strong-field regime is not new (see
Refs.~\onlinecite{Ando,Prasad,Bychkov}). However, its derivation has
remained unsatisfactory: Ando's work~\cite{Ando} is based on the
self-consistent Born approximation (SCBA), which is invalid in this
regime; Prasad and Fujita~\cite{Prasad} ignored the localized nature of
the particle motion; Bychkov and Iordanskii~\cite{Bychkov} {\it de
facto\/} assumed that the electron's trajectory (in the absence of
interactions) is $\delta\omega(\bbox{\rho}) = {\rm const}$ instead of
the correct $U_0 (\bbox{\rho}) = {\rm const}$.

\begin{figure}
\centerline{
\psfig{file=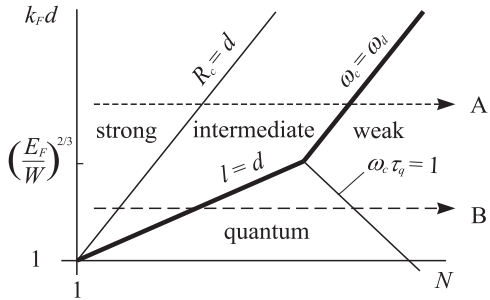,width=2.5in,%
bbllx=215pt,bblly=426pt,bburx=528pt,bbury=620pt}}
\vspace{0.1in}
\setlength{\columnwidth}{3.2in}
\centerline{\caption{
Phase diagram of the CR (double log-scale). $N$ is the index of the
Landau level closest to $E_F$. The bold line marks the location of the
linewidth collapse. The paths A and B, which the 2DEG follows as the
magnetic field decreases, correspond to the classical
(Fig.~\protect\ref{linewidth}) and the quantum
(Fig.~\protect\ref{linewidth_2}) cases, respectively.
\label{diagram}
}}
\end{figure}

So far we have studied the case $k_F d \gg (E_F / W)^{2/3}$. In the
opposite limit an additional ``quantum'' region appears on the phase
diagram of Fig.~\ref{diagram}~\cite{Fogler_98}. Its boundary is formed
by the lines $l = d$ where $l = \sqrt{\hbar / m\omega_c}$ is the
magnetic length, and $\omega_c\tau_q = 1$ where $\tau_q \sim \tau / (k_F
d)^2$ is the quantum lifetime. The lower edge of this region, i.e., the
line $k_F d = 1$ was discussed in the beginning of this Letter. Now we
will show that the collapse of the CR linewidth occurs in the case $1
\ll k_F d \ll (E_F / W)^{2/3}$ as well, and that its position on the
phase diagram is given by the line $l = d$ (which is the same as $R_c /
d = k_F d$). Contrary to the ``classical'' case, the quantum collapse
takes place when the Landau levels are well separated.

%
% FIG. 3
%

\begin{figure}
\centerline{
\psfig{file=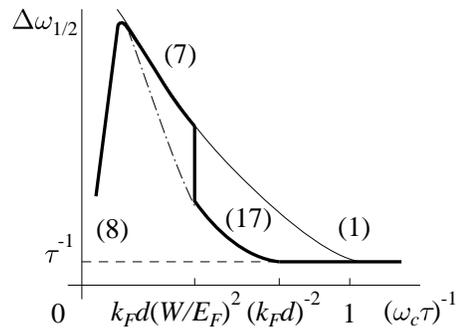,width=2.3in,%
bbllx=170pt,bblly=285pt,bburx=436pt,bbury=480pt}}
\vspace{0.1in}
\setlength{\columnwidth}{3.2in}
\centerline{\caption{
Same as Fig.~\protect\ref{linewidth} but for the path B
of Fig.~\protect\ref{diagram}.
%which corresponds to $1 \ll k_F d \ll (E_F/W)^{2/3}$.
\label{linewidth_2}
}}
\end{figure}

The CR lineshape in the quantum region can be studied within the
SCBA. Doing the integration in Eq.~(2.12) of Ref.~\onlinecite{Ando} for
the case of $E_F$ positioned between the Landau levels and
$|\Delta\omega| \ll \sqrt{\omega_c / \tau_q}$, we find
that the CR line is roughly Lorentzian, while its median width
is given by
\begin{equation}
  \Delta\omega_{1/2} = 0.66\,\tau^{-1} \sqrt{\omega_c \tau_q}.
\label{delta_omega_SCBA}
\end{equation}
On the other hand, in the intermediate-field region $\Delta\omega_{1/2}$
is given by Eq.~(\ref{delta_omega_intermediate}). Therefore, the
crossing of the $l = d$ line causes the drop of $\Delta\omega_{1/2}$ by
a large factor of $k_F d$ (Fig.~\protect\ref{linewidth_2}). The physics
of this ``quantum'' collapse is quite similar to that of the classical
one: the crossing of the line $l = d$ is accompanied by an explosive
growth of the quantum localization length of the states near the Landau
level center (see~\cite{Fogler_98} for discussion). The states with
sufficiently large localization length contribute to $\sigma_+(\omega)$
in the form similar to Eq.~(\ref{Drude}) but with $\tau$ replaced by the
{\it effective\/} transport time. Since the density of states at the
Landau level center is larger than the zero-field density of states by a
factor of $\sqrt{\omega_c\tau_q}$~\cite{Ando}, this effective transport
time is smaller than $\tau$ by the same factor in agreement with
Eq.~(\ref{delta_omega_SCBA}).

%
% FIG. 4
%

Until now we considered $W$ a fixed parameter of the theory. In a more
realistic model (see above) $W$ is determined by the concentration $n_i$
of randomly positioned donors and the screening properties of the 2DEG.
Away from the strong-field regime, the screening is very much the same
as in zero field, and one obtains $W \sim E_F \sqrt{n_i} / (k_F^2 d)$.
Thus, the ``quantum'' case $k_F d \ll (E_F / W)^{2/3}$ is realized if
$n_i$ is sufficiently low, $n_i \ll k_F / d$. On the other hand, {\it
in\/} the strong-field regime (far from the collapse), the amount of
screening, $W$, and $\Delta\omega_{1/2}$ oscillate with the filling
factor~\onlinecite{Filling_factor_oscillations}. A consistent
description of such oscillations must take into account the nonlinear
screening effects~\cite{Fogler_95}.

%Other effects of the electron-electron interactions may influence the
%properties of the CR as well, which requires further study.

In conclusion, let us discuss the experimental side.
%The required range of magnetic fields corresponds to the source signal
%frequencies $\omega / (2 \pi) \sim 100 {\rm GHz}$, which are not readily
%available. It is little wonder that the experimental data are scarce.
The non-monotonic dependence of $\Delta\omega_{1/2}$ on the magnetic
field with the maximum at $R_c \sim d$ has been observed in
Ref.~\onlinecite{Watts}, in agreement with our theory. Arguably, the
collapse of the CR linewidth has also been seen (for the $400{\rm\AA}$
spacer sample). However, a decisive confirmation of the latter prediction
requires further experiments.

M.~M.~F. is supported by DOE Grant No.~DE-FG02-90ER40542 and B.~I.~S. by
NSF Grant No.~DMR-9616880. We thank A.~Yu.~Dobin, M.~I.~Dyakonov,
Yu.~M.~Galperin, and A.~A.~Koulakov for very useful discussions.

%\vspace{-0.2in}

\end{multicols}

\begin{references}
\bibitem{Comment_on_DC} We are talking about $\sigma_{xx}(0)$ on
not too large length scales where the quantum localization effects can
be neglected.
\bibitem{Comment_on_shape} A more careful approach confirms that
$\Delta\omega_{1/2} \sim [\delta\omega]_{\rm rms}$ but refutes the
Gaussian lineshape as an artifact of the $\delta\omega = \text{const}$
approximation.
\bibitem{Comment_on_s} For $H = -1 / 2$ the theory of
Ref.~\onlinecite{Isichenko} gives $s = 4 (1 - H) / (10 - 3 H) =
12 / 23$, so that $e^{2 / (2 s - 1)} \approx 10^{20}$.
\end{references}
\end{document}